\pdfoutput=1
\documentclass{aastex62}

\newcommand{\gjsix}{GJ~674}

\received{}
\revised{}
\accepted{}
\submitjournal{ApJL}

\begin{document}

\title{A Hot Ultraviolet Flare on the M Dwarf Star GJ~674}

\correspondingauthor{Cynthia S. Froning}
\email{cfroning@astro.as.utexas.edu}

\author[0000-0001-8499-2892]{Cynthia S. Froning}
\affiliation{McDonald Observatory, University of Texas at Austin, Austin, TX 78712}

\author{Adam Kowalski}
\affil{Laboratory for Atmospheric and Space Physics, University of Colorado, 600 UCB, Boulder, CO 80309}
\affil{Department of Astrophysical and Space Astronomy, University of Colorado, 389 UCB, Boulder, CO 80309}
\affil{National Solar Observatory, University of Colorado at Boulder, 3665 Discovery Drive, Boulder, CO 80303}

\author[0000-0002-1002-3674]{Kevin France}
\affil{Laboratory for Atmospheric and Space Physics, University of Colorado, 600 UCB, Boulder, CO 80309}

\author{R.~O. Parke Loyd}
\affil{School of March and Space Exploration, Arizona State University, Tempe, AZ 85287}

\author{P.\ Christian Schneider}
\affil{Hamburder Sternwarte, Gojenbergsweg 112, 21029 Hamburg }

\author{Allison Youngblood}
\altaffiliation{NASA Postdoctoral Program Fellow}
\affil{Goddard Space Flight Center, Greenbelt, MD 20771}

\author{David Wilson}
\affil{McDonald Observatory, University of Texas at Austin, Austin, TX 78712}

\author{Alexander Brown, Zachory Berta-Thompson, J. Sebastian Pineda, Jeffrey Linsky}
\affil{University of Colorado, Boulder, CO 80309}

\author{Sarah Rugheimer}
\affil{University of Oxford, Clarendon Laboratory, AOPP, Sherrington Road, Oxford, OX1 3PU, UK}

\author{Yamila Miguel}
\affil{Leiden Observatory, P.O.\ Box 9500, 2300 RA Leiden, The Netherlands}

\begin{abstract}

As part of the Mega-MUSCLES Hubble Space Telescope (HST) Treasury program, we obtained time-series ultraviolet spectroscopy of the M2.5V star, GJ~674. During the FUV monitoring observations, the target exhibited several small flares and one large flare ($E_{FUV} = 10^{30.75}$ ergs) that persisted over the entirety of a HST orbit and had an equivalent duration  $>30,000$~sec, comparable to the highest relative amplitude event previously recorded in the FUV. The flare spectrum exhibited enhanced line emission from chromospheric, transition region, and coronal transitions and a blue FUV continuum with an unprecedented color temperature of $T_{c} \simeq 40,000\pm10,000$~K. In this paper, we compare the flare FUV continuum emission with parameterizations of radiative hydrodynamic model atmospheres of M star flares. We find that the observed flare continuum can be reproduced using flare models but only with the ad hoc addition of hot, dense emitting component. This observation demonstrates that flares with hot FUV continuum temperatures and significant EUV/FUV energy deposition will continue to be of importance to exoplanet atmospheric chemistry and heating even as the host M dwarfs age beyond their most active evolutionary phases.

\end{abstract}


 \section{Introduction} \label{sec:intro}

Active low-mass M (dMe) stars are subject to frequent, energetic stellar flares. The energies can exceed values seen in solar flares by factor of up to 1E6 on the most active M dwarfs \citep{osten2010,osten2016}  and large flares are observed on relatively inactive M dwarfs on occasion as well \citep{paulson2006}.   During the impulsive phase of the flare, the stellar atmosphere is believed to be heated by a beam of mildly relativistic, non-thermal electrons accelerated after magnetic field reconnection and directed from the corona along the field lines into the stellar chromosphere. The electron beam drives a downflowing compression of the chromosphere; these ``chromospheric condensations'' (CCs) are possible sources of the visible band signatures of M star flares. Radiative-hydrodynamic (RHD) simulations of the CCs can match the observed characteristics of dMe flares by enhancing the levels of electron beam heating over solar models, but such models require the introduction of multiple emitting regions on the star, including regions with lower densities, to provide reasonable Balmer line broadening consistent with observations. They also require flux density levels that may drive beam instabilities, suggestig alternate heating mechanisms may be at play \citep{kowalski2015,kowalski2017}.

Scientists investigating flare physics in dMe stars have long pursued dedicated observing programs, but the field is also benefitting from the increased interest in the spectral properties of exoplanet host stars, particularly from the expanding observational sample of ``optically inactive" M dwarf stars. One such survey, the Mega-MUSCLES Treasury Survey, is probing the high energy radiation environment of M stars. Mega-MUSCLES and its predecessor, MUSCLES, use panchromatic (5~\AA\ -- 5.5 $\mu$m) spectroscopy and time-series monitoring of M stars to characterize the stellar energetic radiation environment and activity levels, and their effects on exoplanet atmospheres and habitability \citep{france2013,france2016,youngblood2016,loyd2016,youngblood2017,loyd2018a}. 

As part of the Mega-MUSCLES survey, we observed the M2.5V (0.35~M$_{\odot}$) star, \gjsix. It hosts a $m \sin i = 11.09 M_{\oplus}$ planet in a 4.69 day (0.039 AU) orbit \citep{bonfils2007}. \gjsix\ is classified as a weakly active star: while there is no emission signature in H$\alpha$ characteristic of flare stars, there is an emission reversal in \ion{Ca}{2} and signs of regular spot activity. 
The MUSCLES survey has shown that even optically inactive M stars exhibit frequent ultraviolet flare activity \citep{france2016,loyd2018a}. \gjsix\ is no exception. During five monitoring orbits, Hubble Space Telescope (HST) far-ultraviolet (FUV) observations caught several small flares and one large one. The large flare exhibited strong emission in lines tracing chromospheric and coronal regions of the upper atmosphere of the star and blue FUV continuum emission with a hot ($T_{c} \approx 40,000$~K) color temperature. 

The latter phenomenon is the motivation for this work.  In this manuscript, we present the properties of this unique flare observation and investigate models to reproduce them. In Section~\ref{sec_obs}, we describe the observations and data analysis. Section~\ref{sec_models} presents the RHD model parameterizations and fits to the flare FUV flux. Section~\ref{sec_discussion} discusses the implications of the results for our understanding of flare physics and the effects of such flares on exoplanet atmospheres in M stars.

\section{Observations} \label{sec_obs}

We observed \gjsix\ in 2018 April 2--3 using COS on HST \citep{green2012}. The visit included five HST orbits in COS G130M, providing time-resolved FUV spectroscopy of the star. The COS G130M/1222 mode has spectral coverage from 1065--1365~\AA\ at resolving powers of $R > 10,000$.  The time-series spectral light curves were extracted from the time-tag event files for each exposure by summing the counts around the location of the spectral trace in the 2-D event list and subtracting background regions of the same size offset in the cross-dispersion direction. 

\begin{figure*}
\plottwo{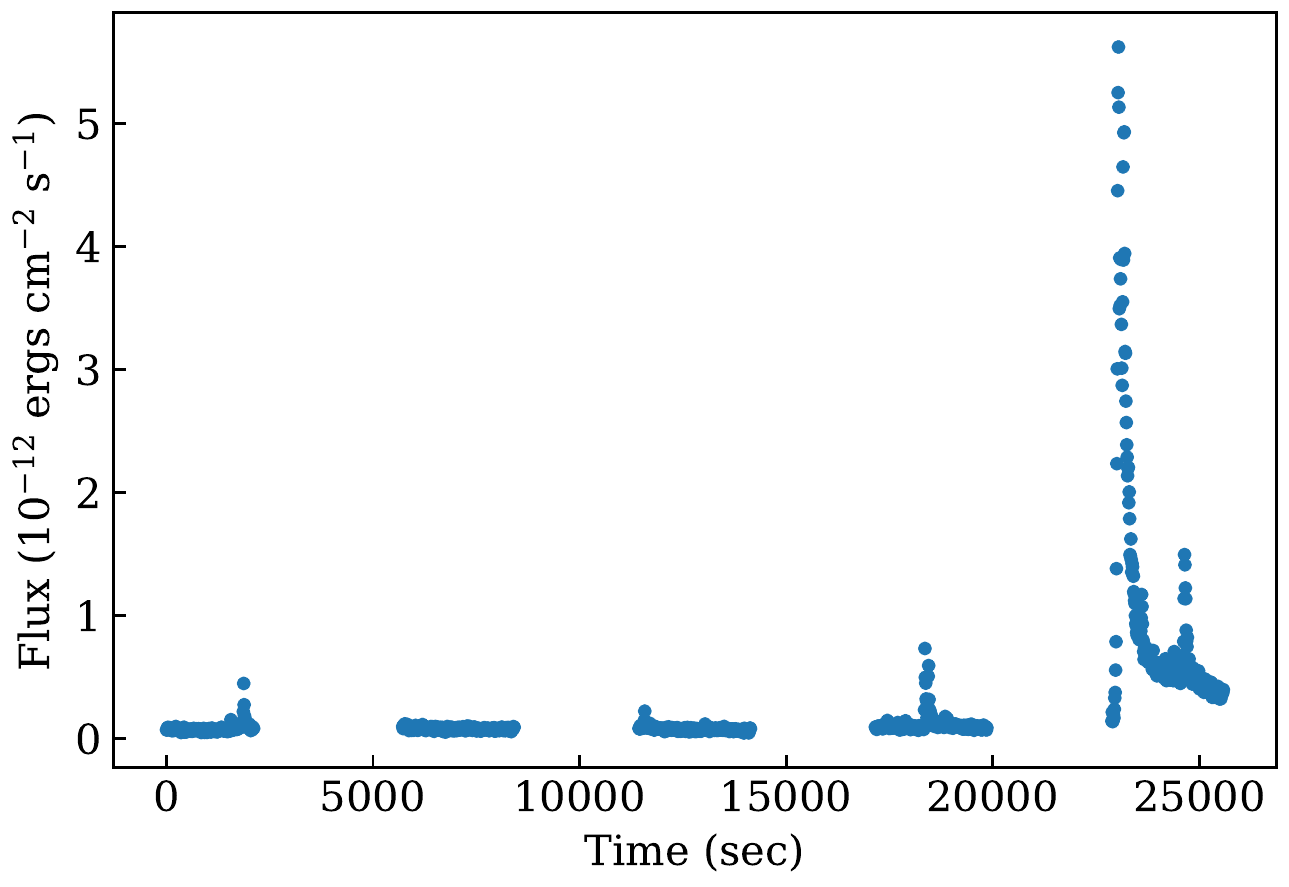}{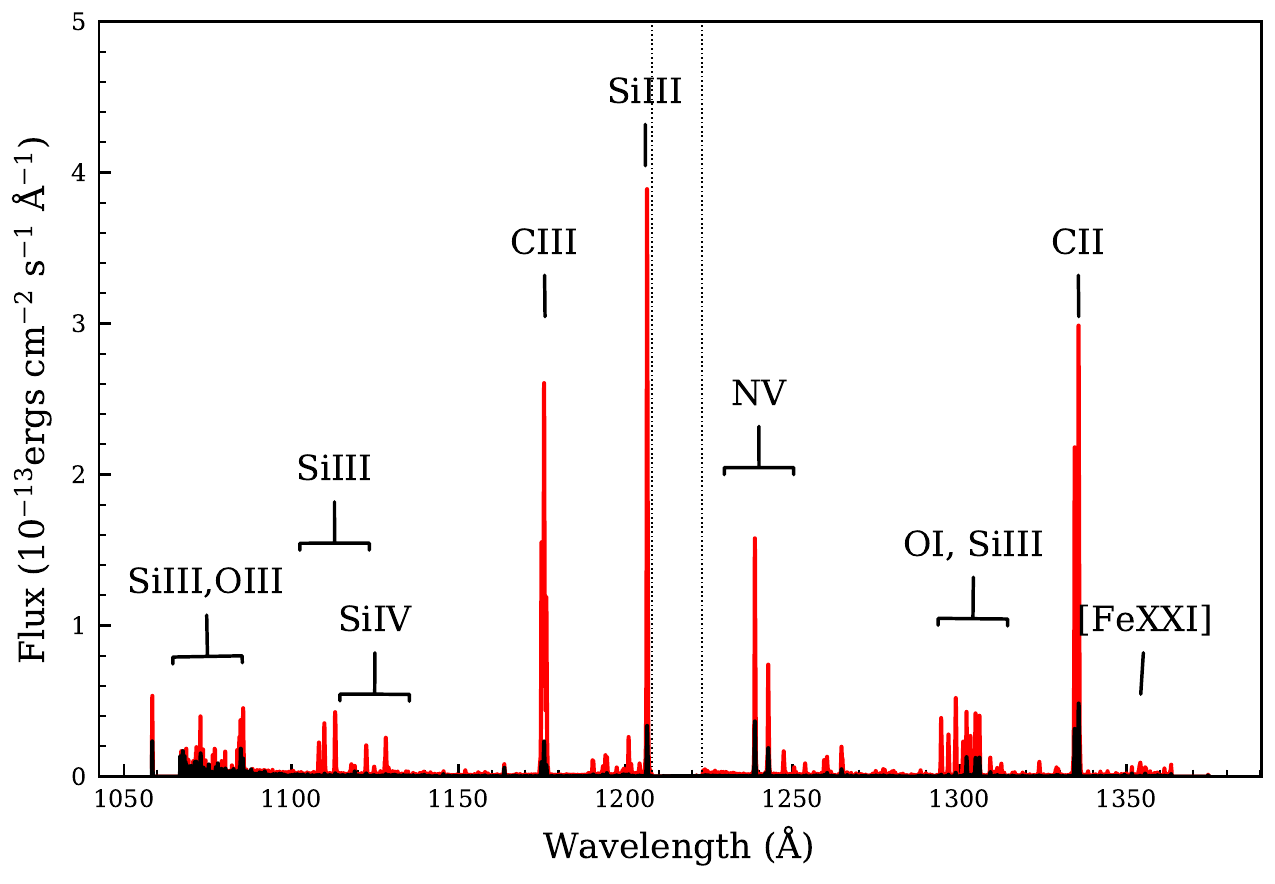}
\caption{The left panel shows the light curve of the COS observations of \gjsix, taken 10 sec sampling intervals. The gaps are HST occultations during each orbit. The right panel compares the flare and out of flare spectra.  The dotted lines in the right panel shows the gap between the two COS detector segments, in which the Ly$\alpha$ line is placed in this grating setting.\label{fig_lcs}}
\end{figure*}

The COS lightcurve and the quiescent and flare spectra are shown in Figure~\ref{fig_lcs}. Several small flares were observed during the visit as well as one large flare in the last orbit. We defined the flare spectrum as the entirety of orbit 5 and the out of flare spectrum as the average of the previous four orbits. The FUV flare energy was calculated as per Equation~1 in \citet{loyd2018a} and was found to be $E_{FUV} = 10^{30.75}$~ergs. If defined in terms of the equivalent duration, $\delta$, that the star would have to emit in quiescence to match the observed flare energies, the flare had $\delta \simeq 30,000$~sec. The equivalent duration of the COS flare is 2.0--4.4$\times$ the largest values seen in previous HST FUV observations of M star flares \citet{loyd2018b,loyd2018a}. More energetic flares (in absolute emission energy) have been observed, but because \gjsix\ isn't a ``flare star'' \citep[according to its lack of H$\alpha$ emission in quiescence;][]{kowalski2009}, the COS flare is substantially more energetic relative to the quiescent flux of the star than previously observed flares. In terms of equilvalent duration, the flare rivals the Great Flare observed on AD~Leo by IUE  \citep[$\delta \geq 40,000$~sec]{hawley1991,loyd2018a}.

\begin{figure*}
\plotone{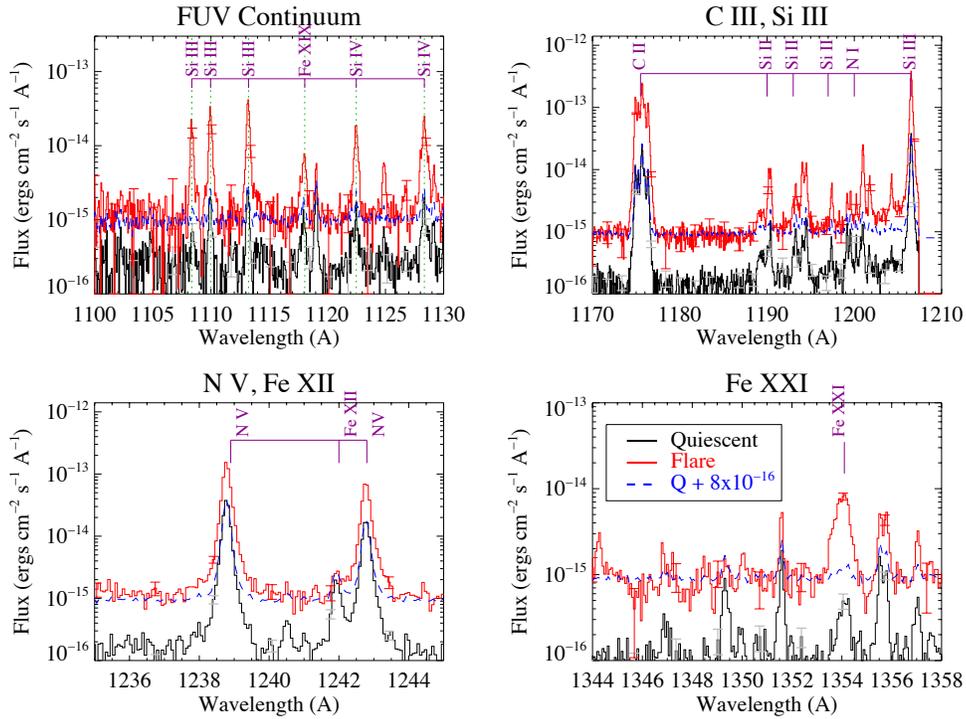}
\caption{An expanded view of several wavelength regions shows the increase in coronal and chromospheric emission lines and the FUV continuum during the flare.  Prominent lines are labeled in both panels.\label{fig_specs}}
\end{figure*}

Figure~\ref{fig_specs} highlights several interesting spectral regions in and out of the flare. During the flare, line emission was enhanced in a number of lines tracing chromospheric and transition region emission (e.g., \ion{C}{2}, \ion{C}{3}, \ion{Si}{3}, \ion{Si}{4}, \ion{N}{5}; with formation temperatures ranging from $T_{form} \sim 3\times10^{4}$~K to $T_{form} \sim 2\times10^{5}$~K) as well as lines tracing hotter coronal regions ([\ion{Fe}{12}], [\ion{Fe}{19}], [\ion{Fe}{21}]; $T_{form} \sim 10^{6}-10^{7}$~K). In general, the lower temperature chromospheric lines showed similar peak to quiescent flux ratios to the largest flare in the MUSCLES sample of optically inactive M stars \citep[on GJ~876;][]{france2016}, with \ion{C}{3} and \ion{Si}{3} increasing in flux by a factor of $\sim100$ and \ion{C}{2} by 30--50. However, the higher formation temperature \ion{N}{5} line and the FUV continuum emission increased by larger factors in the \gjsix\ flare --- $\sim 10$ vs.\ $\sim 5$ for the GJ~876 flare --- and the [\ion{Fe}{21}] $\lambda$1354 line, only marginally detected in the GJ~876 flare, increased by a factor of $\sim$20 here. 

In Table~\ref{tab_fluxes}, we summarize the emission properties of the flare, giving the fluxes, line widths, and velocities for a selection of transitions before the flare and integrated over the flare. We also give the FUV continuum flux at the flare peak, the integrated flare energy in the FUV, and estimates of the integrated EUV flare energy. For the latter, we directly scaled the FUV emission to the EUV using the semi-empirical M dwarf stellar model from \citet{fontenla2016} and also calculated the EUV energy using the relation between the strength of the \ion{N}{5} line and the EUV flux from \citet{france2018}. (Note that both of these estimates rely on quiescent scaling relations applied to the flare spectra.) 

The emission line flare properties are also presented as inputs for future modeling. Here, we note  a few features displayed during the flare. The \ion{N}{5} doublet line ratio indicates that the line is optically thin both in quiescence and during the flare, reflecting its origin in the transition region. The hot coronal transition [\ion{Fe}{21}] exhibits a blue-shift of $\sim -25$~km~s$^{-1}$ (relative to its quiescent velocity) during the flare, indicative of hot coronal plasma being pushed outward by the exploding plasma below.  In the chromospheric lines, the broad components of the lines widen by $\sim 5-25$~km~s$^{-1}$, which may indicate an increase in the turbulent velocity of the line-forming region during the flare. The broad line components also redshift by $\sim5$~km~s$^{-1}$ while the narrow components remain stationary.

\begin{deluxetable*}{ccccccc}
\tablecaption{Line and Continuum Fluxes for the COS Flare\label{tab_fluxes}}
\tabletypesize{\scriptsize}
\tablehead{
\colhead{} & \multicolumn{3}{c}{Quiescent} &  \multicolumn{3}{c}{Flare Peak} \\
\colhead{Line} & \colhead{Flux} & \colhead{Velocity} & \colhead{FWHM} &  \colhead{Flux} & \colhead{Velocity} & \colhead{FWHM} \\
\colhead{} & \colhead{($10^{-15}$ ergs cm$^{-2}$ s$^{-1}$)} &  \colhead{(km s$^{-1}$)} &  \colhead{(km $s^{-1}$)} 
& \colhead{($10^{-15}$ ergs cm$^{-2}$ s$^{-1}$)} &  \colhead{(km s$^{-1}$)} &  \colhead{(km $s^{-1}$)} } 
\startdata
\ion{Si}{3} 1206 n & $6.0\pm0.4$ & $2.3\pm0.5$ & $33\pm2$ & $66\pm1$ & $3.2\pm0.2$ & $37\pm1$ \\
\ion{Si}{3} 1206 b & $4.3\pm0.6$ & $7\pm1$ & $101\pm5$ & $44\pm2$ & $12.4\pm0.5$ & $110\pm1$ \\
\ion{N}{5} 1238 n & $6.0\pm0.4$ & $-10.6\pm0.3$ & 33$\pm$2 & $21\pm2$ & $-11.6\pm0.3$ & 29$\pm$1 \\
\ion{N}{5} 1238 b & $3.0\pm0.9$ & $-11.4\pm0.7$ & $64\pm5$ & $17\pm2$ & $-7.2\pm0.6$ & 70$\pm$2 \\
\ion{N}{5} 1242 n & $2.6\pm0.3$ & $-4.5\pm0.4$ & $32\pm2$ & $14\pm6$ & $-9.1\pm0.3$ & 38$\pm$1 \\
\ion{N}{5} 1242 b & $1.9\pm0.6$ & $-5\pm1$ & $64\pm6$ &  $4.1\pm0.8$ & $3\pm3$ & $97\pm7$ \\
\ion{C}{2} 1334 n & $5.3\pm0.2$ & $-16.4\pm0.2$ & $25\pm1$ &  $32.9\pm0.9$ & $-17.2\pm0.2$ & $28.7\pm0.6$ \\
\ion{C}{2} 1334 b & $2.6\pm0.3$ & $-11.8\pm0.8$ & $75\pm3$ & $23\pm1$ & $-7.3\pm0.6$ & $85\pm2$ \\
\ion{C}{2} 1335 n & $9.1\pm0.4$ & $-15.3\pm0.2$ & $30.8\pm0.9$ & $61\pm1$ & $-17.3\pm0.2$ & $41.7\pm0.6$ \\
\ion{C}{2} 1335 b & $4.6\pm0.6$ & $13.0\pm0.6$ & $75\pm3$ & $28\pm2$ & $-9.8\pm0.6$ & $98\pm2$ \\
{[}\ion{Fe}{12}{]} 1242 & $0.48\pm0.02$ & $-14.9\pm0.6$ & $50\pm2$ & \nodata\tablenotemark{a} & \nodata & \nodata \\
{[}\ion{Fe}{19}{]} 1118 & $0.31\pm0.05$ & $-21\pm4$ & $106\pm12$ & $2.6\pm0.3$ & $-17\pm3$ & $100\pm8$ \\
 {[}\ion{Fe}{21}{]} 1354 & $0.21\pm0.06$ & $12\pm8$ & $101\pm26$ & $4.7\pm0.2$ & $-13\pm2$ & $125\pm4$ \\
\hline
Continuum & Wavelength & & & Flux \& Energy \\
F$_{\lambda}$\tablenotemark{b} & 1142~\AA\ & & & $8.5\pm0.8\times10^{-16}$ ergs~cm$^{-2}$~sec$^{-1}$~\AA$^{-1}$   \\
E$_{FUV}\tablenotemark{c}$ & 1070-1360~\AA\ & & &  $10^{30.75}$~ergs \\
E$_{EUV}\tablenotemark{d}$ & 100-912~\AA\ & &  & $10^{31.5}$~ergs \\
E$_{EUV}\tablenotemark{e}$ & 100-912~\AA\ & &  & $10^{31.8}$~ergs \\
\enddata
\tablecomments{All emission lines, with the exception of the Fe lines (for which the SNR was too low) are fit with double Gaussian profiles with broad (b) and narrow (n) components, consistent with typical fits to M dwarf FUV emission lines \citep{wood1997}. The Fe lines are fit with single Gaussians.}
\tablenotetext{a}{The [\ion{Fe}{12}] line cannot be fit reliably in the flare spectrum because it is blended with the broad wings of Ly$\alpha$. We can place a conservative estimate on the increase in the peak (not integrated) line emission during the flare of $<$50\% relative to its quiescent flux.}
\tablenotetext{b}{The FUV continuum flux density at the peak of the flare.}
\tablenotetext{c}{The integrated flare energy in the FUV, excluding detector gaps and airglow lines, summed over the entirety of orbit 5.}
\tablenotetext{d}{Estimated by scaling the semi-empical M dwarf spectrum of \citet{fontenla2016} to the measured FUV flare energy. }
\tablenotetext{e}{Estimated using the relations between \ion{N}{5} emission and EUV flux from \citet{france2018}. The calculation of the stellar bolometric flux gave $F_{bol} = 2.3\times10^{-8}$~ergs~cm$^{-2}$~s$^{-1}$, assuming $d = 4.5$~pc, $R_{\star} = 0.35 R_{\odot}$, and $T_{eff} = 3400$~K \citep{newton2016,pecault2013}. Then, F(\ion{N}{5}, flare) / F$_{bol} = 2.5\times10^{-6}$ and F(90-911~\AA, flare) = $9.6\times10^{-12}$~ergs~cm$^{-2}$~s$^{-1}$. The overall EUV flare energy, $E(90-911~\AA)$ relies on the \citet{fontenla2016} models to scale the Lyman continuum EUV component to the total EUV emission.}
\end{deluxetable*}

\begin{figure*}
\plottwo{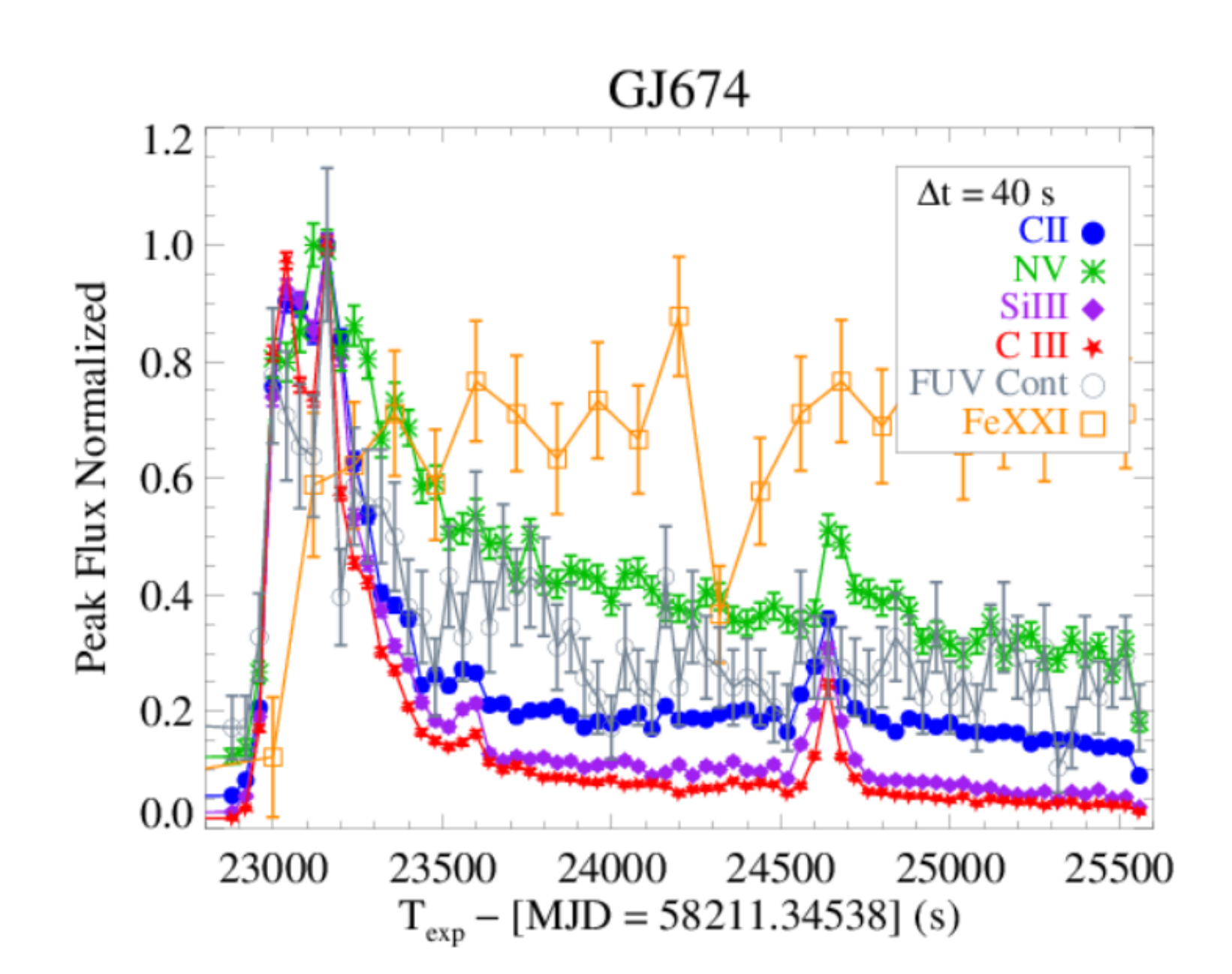}{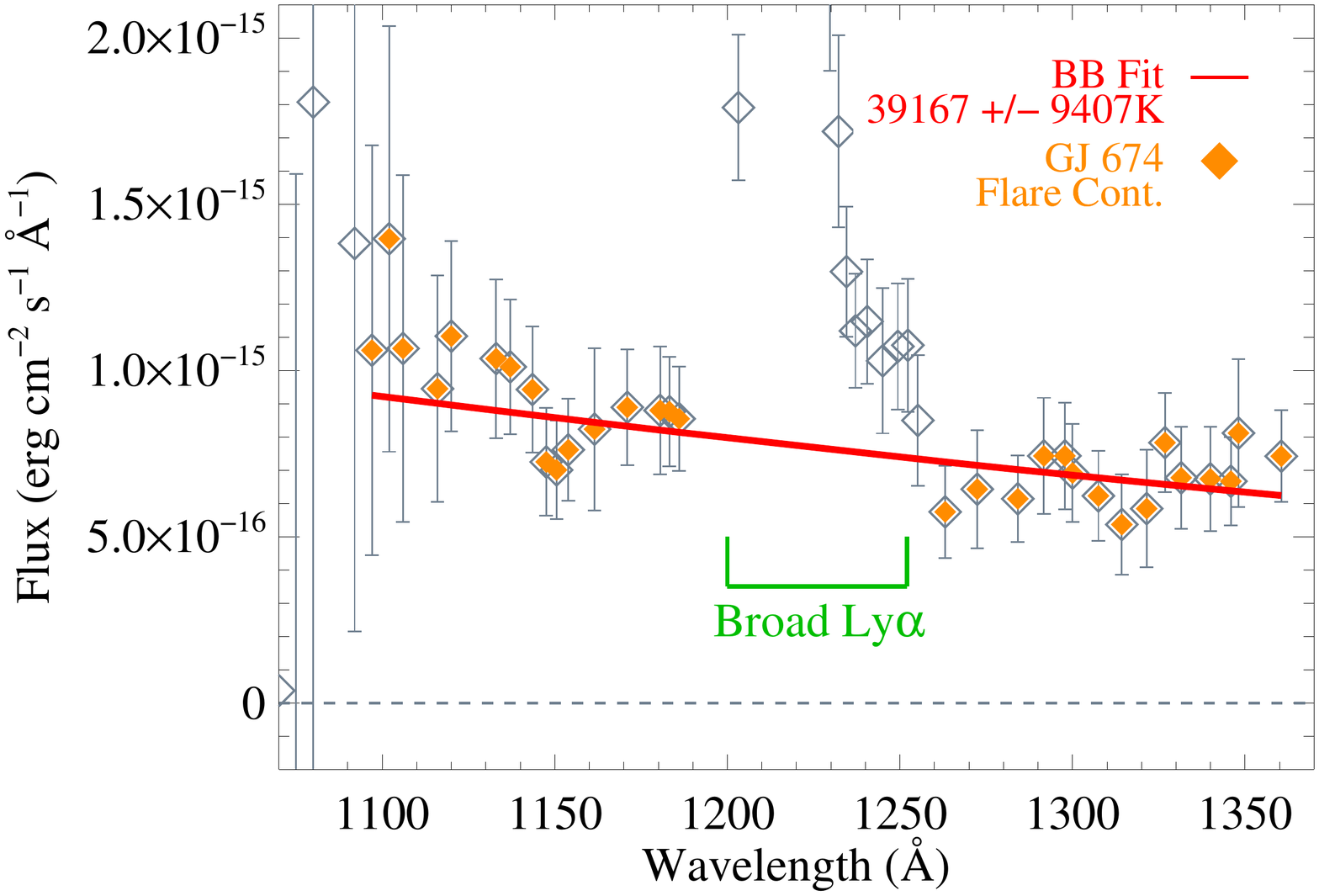}
\caption{The left panel displays the evolution of the COS flare in several emission lines. The lightcurves have been normalized to 1.0 at their peak fluxes. The right panel shows a single temperature blackbody fit to the COS flare spectrum of \gjsix. The method of identifying the continuum and fitting follows that of \citet{france2014}; continuum-only regions were verified against those  identified in \citet{loyd2018a} from high SNR flare spectra. The points in orange were used in the fit, while the broad Ly$\alpha$ emission line wings and the short wavelength end of the waveband where the SNR was low (shown in gray) were excluded. The errors on the data points are the 1$\sigma$ uncertainties in the background-subtracted FUV-continuum counts in 1~\AA\ bins.  \label{fig_bb}}
\end{figure*}

We tracked the time evolution of the emission lines and the continuum during the flare, summarized in the left panel of Fig~\ref{fig_bb}. The light curves have been normalized to the peak flux in each line.  Notably, the FUV continuum traces the time evolution of the chromospheric lines, peaking earlier and declining faster than the coronal [\ion{Fe}{21}] emission. The latter remains high well after the impulsive phase of the flare, indicating ongoing heating in the corona. We fit a single temperature blackbody to the FUV flare continuum emission.  The best fit, shown in Figure~\ref{fig_bb}, has a color temperature of $T_{color} \simeq 40000\pm10000$~K.  It is possible that the FUV continuum contains contributions from unresolved recombination continua and a superposition of discrete emission lines. However, the HST-COS data analyzed here are of sufficient depth and spectral resolution to separate and resolve neutral and singly ionized atomic species.  Additionally, bins that showed evidence of discrete spectral features or RMS flux uncertainty $>$ 50\% of the bin flux were discarded.

\section{Model Fits} \label{sec_models}

RHD models of chromospheres predict the formation of a hot, dense CC layer caused by electron beam heating. When the beam energy flux rate is high, the flaring regions exceed a continuum optical depth of $\tau \sim 1$ and produce a spectrum with a characteristic temperature $T \sim 10^{4}$~K with a small Balmer jump, consistent with the signatures of flares seen in optical observations of dMe stars \citep[e.g.,][]{hawley1992,fuhrmeister2011,kowalski2013,kowalski2016}.
A recent paper, \citet[hereafter K18]{kowalski2018}, has developed parameterizations of CC properties. The parameterizations can be used to predict observables from existing RHD model flare atmospheres without generating computationally intensive full simulations. Using  a suite of existing RHD models, K18 developed prescriptions for the continuum optical depth and emergent flux as a function of $T_{ref}$ and $m_{ref}$, the temperature and column mass ($m$ is in units of g~cm$^{-2}$) at the chromosphere depth in which the chromospheric gas is in steady state with the electron beam heating.

\begin{figure*}
\plotone{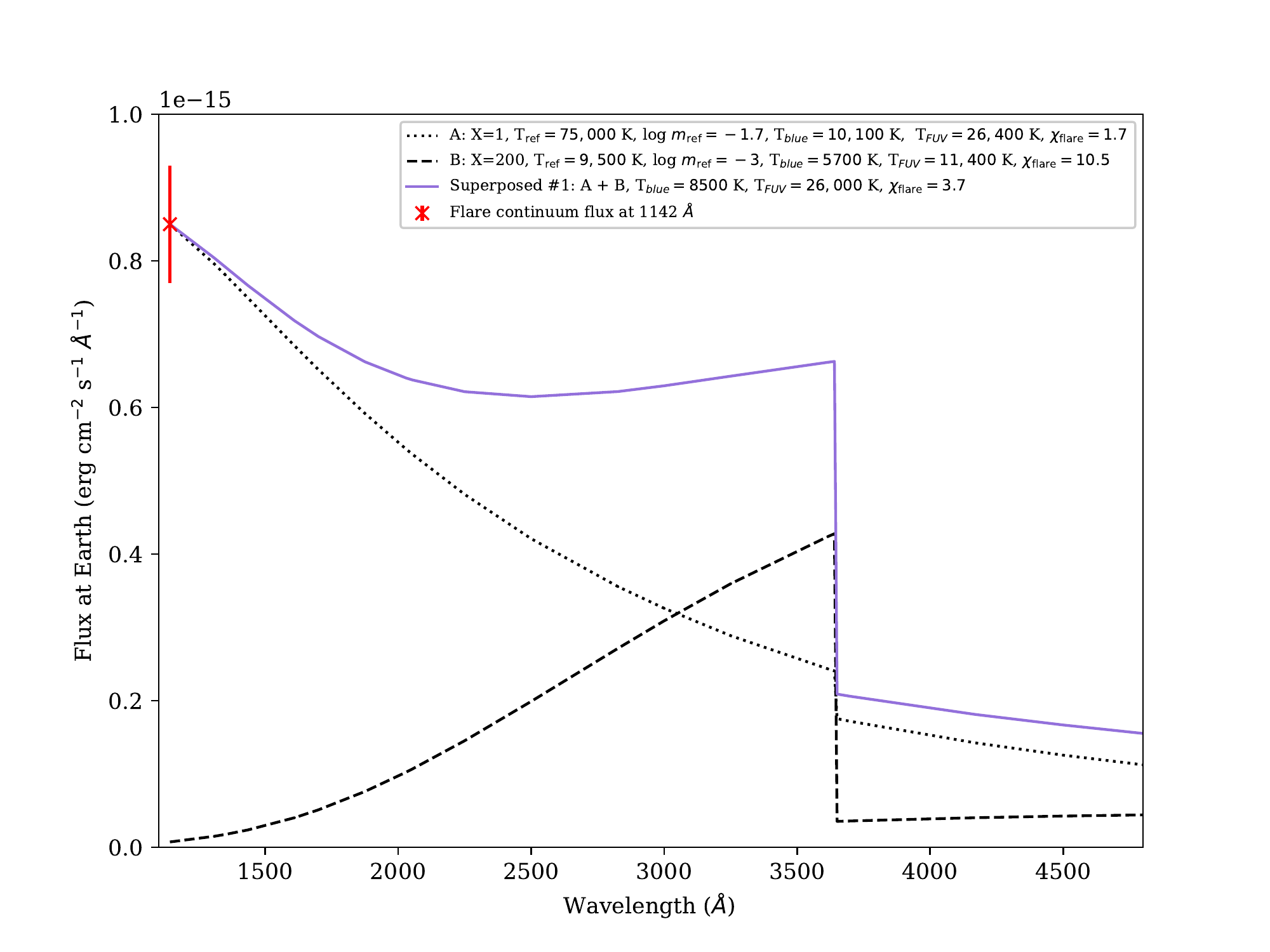}
\caption{Flare model flux densities in the FUV and optical. The dashed line is parameterization of a F11 solar flare model of K18 multiplied by a factor of 200 to match optical color temperature and Balmer jump parameters seen in dMe flares. The dotted line is a $T_{ref} = 75,000$~K, $\log m_{ref} = -1.7$  downflowing compression layer (the stationary flare layers are set to zero). The solid blue line is the superposition of the previous two models. The $\chi_{flare}$ values give the size of the Balmer jump ratio ($\equiv F_{\lambda=3500}/F_{\lambda=4170}$).\label{fig_cc}}
\end{figure*}

We use these parameterizations to investigate modifications to the RHD models needed to fit the hot, blue FUV continuum emission seen in the \gjsix\ flare. RHD models with very high electron beam energy fluxes of $10^{13}$~ergs~cm$^{-2}$~s$^{-1}$ can generate the $T \sim 10^{4}$~K thermal component seen in previously observed flares but not the blue FUV continuum observed here, as the models turn over in the NUV \citep{kowalski2015}. Instead, Figure~\ref{fig_cc} shows an example composite model selected to be consistent with the observed FUV continuum color temperature during the COS flare. The first model (dotted line) is a very hot, dense ($T_{ref} = 75,000$~K, $\log m_{ref} = -1.7$) emitting slab of material, generated from a CC model consisting only of the downflowing condensation layer (emission from the stationary layer has been set to zero).  The second model (dashed line) is a CC parameterization that employs a value of $\log m_{ref} = -3.0$, characteristic of a F11 ($10^{11}$~ergs~cm$^{-2}$~s$^{-1}$) electron beam. It has been scaled up by a factor of 200 such that, when superposed with the hotter component, the combination spectrum exhibits a color temperature in the optical ($\lambda > 4000$~\AA)  of 9000~K and a moderate Balmer jump ratio \citep{kowalski2018b}. Adding the two models (blue curve) results in a FUV color temperature of $T_{C} \simeq 26,000$~K. If the F11 model is not included, the hot component alone produces optical temperatures and Balmer jump ratios ($T_{C} \simeq 12,000$~K and $\chi_{flare} = 1.7$) that are only seen in the most extreme dMe flares and may be unlikely for the less active \gjsix. Thus, while we can match the observed FUV flare continuum, it requires the addition of a hot, dense CC component not predicted by current RHD model simulations.

\section{Discussion} \label{sec_discussion}

With a rotation period of 33.4~d and a likely age of 0.1 to a few Gyr, \gjsix\ sits at an intermediate position between rapidly rotating, young dMe stars ($P_{rot} < 10$~d, ages $<2$~Gyr)  and older, spun down stars \citep[$P_{rot} > 70$~d, ages of several Gyr][]{kiraga2007,bonfils2007,newton2016}. The lack of H$\alpha$ in emission and a $P_{rot} > 30$~d put \gjsix\ on the inactive side of the mass-period relation for a M2.5V, 0.35~$M_{\odot}$ star \citep{newton2018}. In the FUV, however, \gjsix\ was active, a result consistent with the analysis of \citet{loyd2018a} and \citet{loyd2018b}, which demonstrated that, when normalized by the basal flux of the star, dMe and inactive M stars show the same FUV flare energy-frequency relation. The large flare we observed is unprecedented in several respects. In terms of the absolute flare energy, it is not particularly strong compared to historic dMe flare observations. However, the equivalent duration of the flare places it in the realm of the highest amplitude flares observed. Moreover, the color temperature of the continuum, fit with a blackbody of $T_{C} \simeq 40,000$~K, is this hottest spectroscopic color temperature contribution to a broadband flare spectrum measured to date.

While direct constraints from FUV observations of M dwarf flares have been relatively scarce to date, there have been previous indications of the presence of very hot, dense gas emission during flares. The recent large flare captured by the HAZMAT team had a FUV  (1170--1430~\AA) continuum fit by a 15,500~K blackbody; while the continuum emission in that flare was red in the FUV, its color temperature was still higher than those usually seen in optical flare spectra \citep{loyd2018b,kowalski2013}. FUV spectra taken during the AD~Leo Great Flare also showed a flat UV continuum \citep{hawley1991}. Photometric observations of flares in the FUV (1344--1786~\AA) and NUV (1771--2831~\AA) channels with GALEX have also suggested the presence of two distinct flare components, one hotter and of shorter duration, that reaches $T_{C} = 50,000$~K at its peak \citep{robinson2005,welsh2006}. 

Regarding the origin of the FUV continuum, its properties indicate that higher column mass is heated than predicted by any RHD beam model so far explored. Our investigation shows that a combination of a CC and an additional hot chromospheric emitter could replicate the FUV continuum temperature while providing reasonable values for the optical color temperature and Balmer jump ratio. 

A coronal origin, from thermal 30--50~MK gas, may also be possible but generating the requisite gas density is challenging. For example, \citet{caspi2010} also found two components of X-ray continuum emission in an energetic (X4.8) solar flare: a superhot ($T_{e} > 30$~MK) component, likely due to direct coronal heating and compression during the magnetic reconnection process \citep{longcope2011,longcope2016}, and a hot ($T_{e} \lesssim 25$~MK) component that they attributed to chromospheric evaporation. However, an extension of their model spectra to the FUV predicts $F_{1142} = 5\times10^{-17}$~erg~cm$^{-2}$~s$^{-1}$~\AA$^{-1}$, a factor of $\sim$20 below the observed FUV flux. The FUV continuum emission also followed the rapid rise and fast initial decay of the chromospheric lines more closely than the delayed rise and long persistence of the coronal emission.  Further observations of optically inactive M stars in the optical could be used to constrain these models based on optical Balmer jump and color temperature measurements, as well as contraining the presence or absence of multiple emitting components. 

With respect to exoplanets around inactive M stars, the \gjsix\ large flare illustrates the continued impact of stellar irradiance on planetary atmospheres well after the runaway greenhouse phase on the pre-main sequence and the young, active dMe stage of the star.  UV continuum fluxes with color temperatures in excess of 40,000~K during flares  can increase the photolysis rate by 1--2 orders of magnitude over a canonical 9,000~K flare continuum temperature \citep{loyd2018a}. The EUV/FUV energy depositions onto the surface of the planet during the flare has a particular effect on molecules such as H$_{2}$, N$_{2}$, CO, and CO$_{2}$ whose photodissociation cross sections peak $<1200$~\AA, as well as H$_{2}$O in small gas planets \citep{loyd2016,miguel2015,venot2016}. Thus, the impacts of the high UV flux emitted during flares on the chemistry and potential habitability of M star exoplanets will continue to be of importance even as the stars age beyond their most active evolutionary phases.

\section{Conclusions}

We observed the M2.5V exoplanet host star, \gjsix, using COS on HST. During the FUV spectroscopic observations, the star had several small flares and one large one. The large flare spanned more than a HST orbit, had a FUV flare energy $E_{FUV} > 10^{30.75}$~ergs, and a flare equivalent duration of $\delta > 30,000$~sec, making the energy of the flare relative to the quiescent stellar flux comparable to the most energetic flare previously observed in the FUV. The flare spectrum displayed enhanced emission in lines tracing chromospheric and coronal temperature regimes and a strong, blue continuum with a color temperature of $\simeq$40,000~K,the hottest spectroscopic color temperature contribution to broadband observations of M star flares to date. We investigated modifications to existing RHD flare models needed to match the observed FUV continuum. We find that the addition of a very hot compression slab in the CC region can generate the requisite FUV color temperatures but require gas column masses significantly higher than are produced in current RHD models of stellar flares.Finally, we note the effects of flares with strong, hot FUV/EUV continuum emission on the atmospheres of exoplanets around M stars, particularly the fact that stars that have evolved past the pre-main sequence and young, active stellar phases can still drive significant atmospheric heating and photolysis in exoplanet atmospheres via continued energetic flares.

\acknowledgments

Based on observations made with the NASA/ESA Hubble Space Telescope, obtained from the Data Archive at the Space Telescope Science Institute, which is operated by the Association of Universities for Research in Astronomy, Inc., under NASA contract NAS 5-26555. These observations are associated with program \# 15071. Support for program \#15071 was provided by NASA through a grant from the Space Telescope Science Institute, which is operated by the Association of Universities for Research in Astronomy, Inc., under NASA contract NAS 5-26555. All of the data presented in this paper were obtained from the Mikulski Archive for Space Telescopes (MAST).
 
\vspace{5mm}
\facilities{HST(COS)}

\software{Astropy \citep{2018arXiv180102634T} , Jupyter \citep{kluyver2016}, Matplotlib \citep{hunter2007}, NumPy \citep{vanderwalt2011}}

\end{document}